\documentclass{article}
\usepackage{spconf,amsmath,graphicx,hyperref}
\usepackage{amsfonts}
\usepackage{bbm}
\usepackage{booktabs}
\usepackage{arydshln}

\title{WST-Graph: Topology-Preserving Wavelet Scattering Front-End for Speech Deepfake Detection}
\name{Kwok-Ho Ng, Tingting Song, Bingwen Feng, Zhihua Xia}

\address{College of Cyber Security, Jinan University, Guangzhou, China}
\begin{document}
\ninept
\maketitle
\begin{abstract}
The acoustic front-end determines which forensic cues a speech deepfake detector can exploit. The wavelet scattering transform (WST) provides stable multiscale coefficients with explicit coordinates, yet direct flattening obscures the parent relation between paths. We introduce WST-Graph, reconstructing these paths as a sparse modulation-carrier grid for an AASIST graph backend. Modulation-level normalization and length-aware adaptive local attention pooling produce fixed relative-time representations while retaining the acoustic axes before learned adaptation. This yields a waveform-to-graph interface with a fixed, parameter-free WST. 
Our configurations remain competitive with AASIST while using approximately 60\% fewer trainable parameters and show clear gains on selected out-of-domain benchmarks.
These results underscore the value of preserving parent-child relations within the carrier--modulation topology when constructing a compact, physically grounded interface for graph-based speech deepfake detection. Code will be released at \href{https://github.com/saki-ciallo/wst-graph}{GitHub}.

\end{abstract}
\section{Introduction}
\label{sec:intro}

Recent speech synthesis and voice conversion systems generate increasingly natural speech \cite{zhang2025minimax, lian2026dots, xiang2026qwen}, 
raising risks of impersonation, fraud, and disinformation \cite{masood2023deepfakes} 
and motivating evaluation campaigns such as ASVspoof 
\cite{nautsch2021asvspoof, yamagishi21_asvspoof, wang24_asvspoof}. Speech deepfake detectors (SSD) can also degrade substantially under unseen generators and recording conditions \cite{muller22_interspeech}. Some recent state-of-the-art systems employ self-supervised learning (SSL) front-ends to improve detection performance, but do so at the cost of substantially larger parameter counts \cite{tak22_odyssey}. Moreover, analyzing synthetic artifacts from SSL hidden states is challenging because they lack explicit acoustic coordinates.

Conventional anti-spoofing methods use inspectable magnitude, cepstral, and phase-derived features, including linear frequency cepstral coefficients (LFCC), constant Q cepstral coefficients (CQCC), and modified group delay \cite{yi2023audio, tahaoglu2025deepfake}. Raw waveform-based systems avoid prescribing a fixed hand-crafted spectrum. For example, the end-to-end system RawNet2 jointly learned its waveform encoder with the classifier \cite{tak2021end}. The AASIST applied a parameterized SincNet \cite{ravanelli2018speaker} and a residual convolutional encoder before constructing spectral and temporal graphs \cite{jung2022aasist}. Its Sinc filters retain interpretable passbands, but the subsequent 2-D pooling and residual convolutions mix neighboring frequency-time responses and progressively reduce temporal resolution before graph construction. The resulting channels no longer carry explicit carrier or modulation-frequency coordinates. This motivates a front-end that delays feature fusion across distinct spatiotemporal pathways, thereby preserving explicit acoustic axes for graph processing and coordinate-aligned analysis.

To address this, the wavelet scattering transform (WST) presents a promising candidate \cite{mallat2012group, anden2014deep}. It cascades multiresolution wavelet filterbanks, complex modulus operators, and low-pass averaging filters to produce stable, locally translation-invariant multiscale coefficients. Specifically, first-order coefficients capture the averaged wavelet-modulus responses within carrier bands, while second-order coefficients characterize the temporal modulations of the corresponding first-order envelopes \cite{anden2014deep, anden2015joint}. Crucially, because every second-order path remains explicitly linked to its parent carrier band, the WST inherently preserves a structured carrier-modulation organization before any learned channel mixing.

Recently, WST-X integrated scattering features with SSL representations \cite{xuan2026wst}. Specifically, its 1-D variant (WST-X1) globally pools 1-D scattering coefficients, whereas its 2-D variant (WST-X2) spatially flattens a 2-D scattering tensor computed from SSL feature maps. In contrast, we retain the internal topology of 1-D scattering to serve as an explicit, structured interface for graph-based backends.

To this end, we propose the WST-Graph front-end. First, utilizing the parent relation inherent in WST metadata, we group second-order paths by modulation band and align them according to their parent carrier bands. This constructs a sparse joint modulation-carrier grid without prematurely averaging away valid second-order coefficients. Within this grid, first-order coefficients occupy a dedicated channel, and a fixed binary mask identifies structurally absent path combinations. Second, modulation-level normalization standardizes paths sharing the same modulation band. Within this stage, length-aware channel-wise adaptive local attention pooling (ALAP) segments each valid temporal trajectory into a fixed number of relative-time intervals, pooling every channel-carrier pair independently within each interval. A pointwise adapter then introduces the initial cross-order and cross-modulation fusion exclusively at the same carrier-time coordinate. Following this, three depthwise-separable residual blocks model local carrier-time contexts while preserving the latent grid dimensions, ensuring compatibility with AASIST-style graph backends. We deploy the spectro-temporal graph architecture of AASIST as a concrete backend to model global interactions between the carrier and relative-time nodes. Third, we conduct a three-seed empirical study of the scattering scale and first-order filterbank resolution, together with a controlled comparison of second-order modulation-filter resolution.


\section{Proposed Methods}

\begin{figure*}[h]
    \centering
    \includegraphics[width=0.8\textwidth]{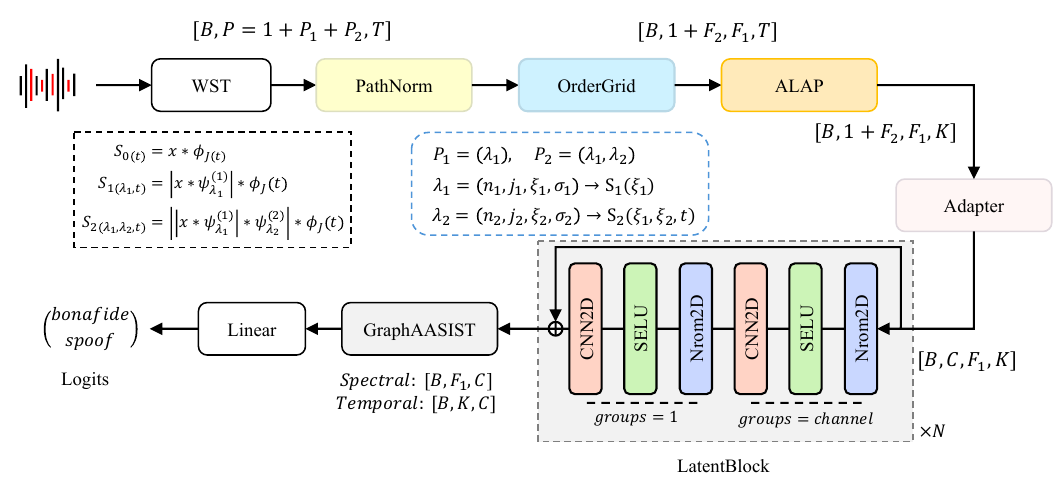}
    \caption{Overview of the proposed WST-Graph pipeline. GraphAASIST follows the graph backbone of the original AASIST architecture.}
    \label{fig:pipeline}
    \vspace{-1.5em}
\end{figure*}


\subsection{Scattering Paths and Path Normalization}
Let $x_b(t)$ denote the $b$-th waveform. The first two orders of a one-dimensional WST are $S_b^{(1)}(\lambda_1,t) =(|x_b*\psi_{\lambda_1}|*\phi_J)(t)$ and $S_b^{(2)}(\lambda_1,\lambda_2,t) =(||x_b*\psi_{\lambda_1}|*\psi_{\lambda_2}|*\phi_J)(t)$, obtained by cascading complex Morlet convolutions, modulus nonlinearities, and low-pass averaging. Here, $\lambda_i$ indexes the wavelet applied at layer $i$. Kymatio \cite{andreux2020kymatio} provides its discrete index $n_i$, scale $j_i$, and normalized center-frequency coordinate $\xi_i$. Thus, a first-order path carries $(n_1,j_1,\xi_1)$, while a second-order key $(n_1,n_2)$ maps to $(\xi_1,\xi_2)$, where $\xi_1$ identifies the parent carrier and $\xi_2$ its envelope-modulation frequency.

We discard the zeroth-order low-pass component and stack all admissible first- and second-order coefficients along Kymatio's path axis. This gives a flat-path tensor $\mathbf{S}\in\mathbb{R}^{B\times P\times T}$, where $P=P_1+P_2$, and $P_1$ and $P_2$ denote the numbers of retained first- and second-order paths, respectively. Before OrderGrid recovers the carrier--modulation topology described above, PathNorm maps the non-negative scattering amplitudes to the log domain to compress their dynamic range,
\begin{equation}
    X_{b,p,t}=\log\left(S_{b,p,t}+\epsilon\right). 
    \label{eq:pathnorm}
\end{equation}

Let $L_b\leq T$ be the effective number of valid WST frames for the $b$-th training utterance, excluding temporal padding. For a designated statistics aggregation group $g$, we define its valid support domain $\Omega_g=\{(b,p,t):g(p)=g,\ 0\leq t<L_b\}$ and the corresponding sample count $N_g=|\Omega_g|$. PathNorm estimates the group-specific mean and standard deviation, respectively:
\begin{align}
    \mu_g&=\frac{1}{N_g}\sum_{(b,p,t)\in\Omega_g}X_{b,p,t}, \\
    \sigma_g&=\sqrt{\max\left(\frac{1}{N_g}\sum_{(b,p,t)\in\Omega_g}X_{b,p,t}^{\,2}-\mu_g^2,0\right)}.
    \label{eq:mu_sigma}
\end{align}
The normalized scattering paths are subsequently obtained by: 
\begin{equation}
    Z_{b,p,t}=\frac{X_{b,p,t}-\mu_{g(p)}}{\max(\sigma_{g(p)},\sigma_{\min})+\epsilon},
    \label{eq:norm_overview}
\end{equation}
where $\sigma_{\min}$ enforces a strict lower bound on the scaling factor. We consider three distinct group assignment policies: 
\begin{equation}
    {log\_path}(g_\mathrm{path}(p)=p),\ {log\_order}(g_{\mathrm{order}}(p)=o(p)),
\end{equation}
and ${log\_modulation}$, defined as:
\begin{equation}
    g_{\mathrm{mod}}(p)=
    \begin{cases}
        0,&o(p)=1,\\
        \mathrm{grp}(n_2(p)), &o(p)=2.
    \end{cases} 
\end{equation}

The $log\_modulation$ policy groups first-order paths together and second-order paths by $n_2$, standardizing each group while preserving its carrier-dependent variation. We also evaluate two baselines: $log\_only$, which bypasses standardization ($\mathbf{Z}=\mathbf{X}$), and $none$, which further omits the log transform ($\mathbf{Z}=\mathbf{S}$). Crucially, the effective length $L_b$ isolates padded frames exclusively during training statistics estimation; a subsequent temporal mask prevents them from contaminating the ALAP operation. Because structurally absent cells are introduced only during grid mapping (OrderGrid), a topology mask is omitted at this stage, and every policy strictly preserves the $B\times P\times T$ tensor shape.

\subsection{Topology Recovery via OrderGrid}

The joint second-order scattering transform is not a full Cartesian product of all first- and second-order wavelets. For the temporal scattering implementation considered here, the Kymatio retains an ordered scattering path if and only if the dyadic scales satisfy $j_1(n_1)<j_2(n_2)$. Consequently, the admissible path set $\mathcal{A}=\{(n_1,n_2):j_1(n_1)<j_2(n_2)\}$ is inherently sparse. For every retained path, Kymatio metadata supplies its scattering order, discrete filter keys, dyadic scale, and center-frequency coordinate. To robustly reconstruct the topology, OrderGrid aligns paths using these discrete keys. Let $f\in\{1,\ldots,F_1\}$ index the first-order keys $n_1(f)$ sorted by increasing center frequency $\xi_1$, and let $m\in\{1,\ldots,F_2\}$ index the unique second-order keys $n_2(m)$ sorted by increasing $\xi_2$. 
Let $\pi_1(f)$ and the partial map $\pi_2(m,f)$ retrieve the corresponding indices on the flat path axis $\{1,\dots,P_1\}$, with $\pi_2(m,f)$ defined only when $(n_1(f),n_2(m))\in\mathcal{A}$.

The normalized scattering paths are subsequently embedded into a structured joint modulation-carrier tensor $\mathbf{G}\in \mathbb{R}^{B\times(1+F_2)\times F_1\times T}$. The grid assignment is defined by $G_{b,0,f,t}=Z_{b,\pi_1(f),t},$ and for the second-order paths:
\begin{equation}
    G_{b,m,f,t}=
    \begin{cases}
        Z_{b,\pi_2(m,f),t}, & \mathrm{if}(n_1(f),n_2(m))\in\mathcal{A},\\
        0, & \mathrm{otherwise},
    \end{cases}
    \ 1\leq m\leq F_2.
    \label{eq:grid1}
\end{equation}
By framing the tensor construction this way, the first-order representation of shape $B\times F_1\times T$ is concatenated along the channel dimension as channel 0, while channels $1\leq m \leq F_2$ contain the available normalized second-order paths for each modulation band $\xi_{2,m}$. This tensor expansion yields a total shape $B\times(1+F_2)\times F_1\times T$, ensuring that no valid scattering coefficient is averaged or discarded prematurely. 
To preserve this underlying structural sparsity, we maintain a fixed structural binary mask $\mathbf{M}\in\{0,1\}^{(1+F_2)\times F_1}$, where $M_{0,f}=1$ and $M_{m,f}=\mathbbm{1} \left[(n_1(f),n_2(m))\in\mathcal{A}\right]$ for $m \geq 1$. Thus, $M_{m,f}=0$ denotes a structural vacancy dictated by the scattering topology rather than a measured zero amplitude response. OrderGrid initializes these cells to zero, and the downstream ALAP leverages $M$ to exclude them from pooling calculations. This fixed structural mask operates independently of the utterance-dependent valid temporal duration $L_b$.

\subsection{Length-aware Channel-wise Temporal Pooling}

All waveforms are padded or cropped to $N$ samples, while $N_b$ records the valid sample count. We obtain $L_b$ by counting WST frame centers that precede $N_b$. The valid prefix interval $[0,L_b)$ is dynamically partitioned into $K$ relative-time intervals:
\begin{equation}
    I_{b,k}=\left[\left\lfloor\frac{kL_b}{K}\right\rfloor,
    \left\lfloor\frac{(k+1)L_b}{K}\right\rfloor\right), \ 0 \le k < K.
    \label{eq:kbins}
\end{equation}
To aggregate temporal features without destroying structural axes, we propose a channel-wise ALAP layer. A $1\times k_t$ depthwise temporal convolution sweeps the frame axis to generate logits $q_{b,c,f,t}$. Crucially, this scorer operates independently across channels $c \in \{0, \ldots, F_2\}$ and is shared across carriers $f \in \{1, \ldots, F_1\}$, avoiding premature feature mixing. For any valid cell where $M_{c,f}=1$, the pooled representation $Y_{b,c,f,k}$ is computed via localized softmax-normalized weights:
\begin{align}
    \alpha_{b,c,f,k,t}&=\frac{\exp(q_{b,c,f,t}/\tau)}{\sum_{u\in I_{b,k}}\exp(q_{b,c,f,u}/\tau)}, \ t\in I_{b,k},\\
    Y_{b,c,f,k}&=\sum_{t\in I_{b,k}}\alpha_{b,c,f,k,t}G_{b,c,f,t},
\end{align}
where $\tau$ is a fixed temperature scaling factor. Cells in structurally absent regions ($M_{c,f}=0$) remain strictly zero-initialized and are excluded from pooling. This operation contracts the variable temporal dimension to a fixed grid size $K$, yielding $\mathbf{Y}\in\mathbb{R}^{B\times(1+F_2)\times F_1\times K}$.

\subsection{Channel Adaptation and Graph Classification}

A pointwise channel Adapter performs the first learned mixing across the $1+F_2$ scattering channels at each carrier--time coordinate $(f,k)$, applying $\mathbf{W}_{\mathrm{ad}}\in\mathbb{R}^{C\times(1+F_2)}$ to map them to $C$ latent dimensions via $\mathbf{H}_{b,:,f,k}=\mathrm{SELU}\left(\mathrm{BN}\left(\mathbf{W}_{\mathrm{ad}}\mathbf{Y}_{b,:,f,k}\right)\right)$, yielding $\mathbf H\in\mathbb{R}^{B\times C\times F_1\times K}$. 
A subsequent latent block is repeated $N$ times (see Fig. \ref{fig:pipeline}) under two spatial convolution configurations: latent-D employs a depthwise spatial convolution, while latent-G utilizes a grouped spatial convolution. Both variants append a full $1\times1$ projection, strictly preserve the dimensions $B\times C\times F_1\times K$, and support ordered D/G compositions.

GraphAASIST forms spectral nodes by reducing the $K$ axis and temporal nodes by reducing the $F_1$ axis $\mathcal{U}^S_{b,f}=\{\mathbf{H}_{b,:,f,k}\}_{k=1}^{K}$, and $\mathcal{U}^T_{b,k}=\{\mathbf{H}_{b,:,f,k}\}_{f=1}^{F_1}$. For either sequence $\mathcal{U}=\{\mathbf{u}_i\}_{i=1}^{R}$, define $\mathbf{A}=\max_i|\mathbf{u}_i|$, $\mathbf{M}=\frac{1}{R}\sum_i\mathbf{u}_i$, $\mathbf{G}_p=\left(\frac{1}{R}\sum_i\max(|\mathbf{u}_i|,\epsilon)^p\right)^{1/p}$, $\mathbf{D}=\mathrm{Std}_i(\mathbf{u}_i)$, $\mathbf{M}_\alpha=\sum_i\alpha_i\mathbf{u}_i$, $\mathbf{D}_\alpha=\mathrm{Std}_\alpha(\mathbf u_i)$, where $\alpha=\mathrm{softmax}(e)$. Let $\Pi(\mathbf{a},\mathbf{b})=\mathbf{W}[\mathbf{a}\Vert\mathbf{b}]$, where $[\cdot\Vert\cdot]$ combines two $C$-dimensional statistics along the feature axis and $\mathbf{W}\in\mathbb{R}^{C\times2C}$. We compare 
$\text{max}(\mathbf{A})$,
$\text{max\_mean}(\Pi(\mathbf{A},\mathbf{M}))$,
$\text{gem\_mean}(\Pi(\mathbf{G}_p,\mathbf{M}))$,
$\text{mean\_std}(\Pi(\mathbf{M},\mathbf{D}))$,
$\text{atten}(\Pi(\mathbf{M}_\alpha,\mathbf{D}_\alpha))$,
$\text{max\_atten}(\Pi(\mathbf{A},\mathbf{M}_\alpha))$.
These operators are applied independently to $\mathcal{U}^S$ and $\mathcal{U}^T$. Asym-A uses $(\Pi(\mathbf{G}_p,\mathbf M),\Pi(\mathbf{A},\mathbf M))$ for the spectral and temporal branches, respectively; Asym-B reverses them. This gives $\mathbf V^S\in\mathbb{R}^{B\times F_1\times C}$ and $\mathbf{V}^T\in\mathbb{R}^{B\times K\times C}$. Attentive pooling uses relative ALAP-bin positions for $\mathcal{U}^S$ and normalized log-frequency positions for $\mathcal{U}^T$; a separate $\log_2\xi_1$ embedding is added to $\mathbf{V}^S$. The nodes enter the homo and hetero graph attention stages of AASIST. Finally, a linear classifier outputs logits.

\section{Experiments and Results}
\subsection{Implementation Details}
\textbf{Datasets and Metrics.} Models are trained on ASVspoof 2019 LA (ASV19), resampled to 16 kHz. Training and development utterances use random 4s crops (64,000 samples) with short files right-padded with zeros; evaluation windows start at the absolute onset. Cross-dataset generalization is benchmarked via Speech DF Arena \cite{11345101}, including 14 distinct sets (13 out-of-domain (OOD)). We report the equal error rate (EER [\%] $\downarrow$) following standard ASVspoof protocols. 
\textbf{Model and Training Setup.} The WST front-end uses $J=8$, $Q=(Q_1,Q_2)=(8,1)$, maximum order 2, and oversampling 1. We set ALAP $K=64$ ($k_t=5$), adapter width $C=64$, and GraphAASIST uses dimensions (64, 32) with max node construction. All networks are optimized for 12 epochs across three seeds using focal loss ($\gamma=2$, weights $[0.9, 0.1]$) and AdamW (learning rate $10^{-3}$ with cosine decay) under FP32/BF16 mixed precision.

\subsection{Temporal Reduction and Local Context}
\textbf{Initial Configuration.} We establish a baseline front-end utilizing a fixed WST, OrderGrid, a pointwise adapter, and the default AbsMax GraphAASIST backend, omitting the latent blocks and fixing PathNorm to $log\_only$ in Eq. (\ref{eq:pathnorm}).
\begin{table}[h]
    \vspace{-1.7em}
    \centering
    \small
    \caption{
        Uniform bin averaging versus ALAP at different temporal resolutions. Results are reported as three-seed average EER (\%) with the best result in brackets. Bold indicates best results.
    }
    \setlength{\tabcolsep}{4pt}
    \begin{tabular}{ccccc}
        \toprule
        Setting & Param & $K=32$ & $K=48$ & $K=64$ \\
        \midrule
        Uniform & 86,024 & 21.81 (18.48)  & \textbf{19.25} (16.17)  & 22.12 (18.49) \\
        ALAP    & 86,073 & 21.46 (19.91)  & 22.05 (19.38)  & \textbf{19.14} (14.91) \\
        \bottomrule
    \end{tabular}
    \vspace{-0.7em}
    \label{tab:Q1a}
\end{table}

Both ALAP and its parameter-free baseline must downsample the valid duration of each channel-carrier trajectory into a fixed number of relative-time bins $K$. The baseline replaces ALAP's learned attention weights with a uniform average, defined as $Y^{\mathrm{avg}}_{b,c,f,k} = \frac{1}{|I_{b,k}|}\sum_{t\in I_{b,k}}G_{b,c,f,t}$. 
We hypothesize that uniform averaging dilutes localized spoofing artifacts, whereas ALAP's channel-wise attention preserves them with negligible parameter overhead. We evaluate $K\in\{32,48,64\}$ in Table \ref{tab:Q1a}. 
ALAP introduces only 49 parameters. While uniform averaging peaks at $K=48$ (19.25\% average, 16.17\% best EER), ALAP demonstrates its advantage at $K=64$ by delivering a lower average EER of 19.14\% and an overall best EER of 14.91\%. Consequently, we provisionally retain ALAP with $K=64$.
\begin{table}[h]
    \centering
    \small
    \caption{
        Results are reported in EER (\%), with configurations M1 (DG), M2 (GD), M3 (DDG), and M4 (DGG).
    }
    \setlength{\tabcolsep}{3.8pt}
    \begin{tabular}{ccc|cc|c}
        \toprule
        $N$ & Param & Type D & Param & Type G  & Mixed \\
        \midrule
        1 & 91k  & 8.38 (6.15)  & 95k   & 8.21 (8.10) & M1: 6.74 (5.68) \\
        2 & 96k  & 6.52 (5.77)  & 104k  & \textbf{6.80} (6.32) & M2: 5.92 (4.22) \\
        3 & 100k & \textbf{4.98} (4.80)  & 113k  & 7.03 (5.88) & M3: \textbf{5.46} (5.01)\\
        4 & 105k & 5.74 (4.77)  & 122k  & 6.82 (5.64) & M4: 6.32 (4.88)\\
        \bottomrule
    \end{tabular}
    \vspace{-1.7em}
    \label{tab:Q1b}
\end{table}

Based on this choice, we investigate whether the local carrier-time context should be modeled using depthwise (D), grouped (G), or mixed residual blocks across different block counts $N$. As shown in Table \ref{tab:Q1b}, the D family performs optimally at $N=3$ (D3), achieving the lowest average EER of $4.98\%$ with 100k parameters. Increasing the depth to $N=4$ or introducing grouped interactions leads to performance degradation. Consequently, D3 is retained for the following task.

\subsection{Normalization and Graph-Node Construction} 
With ALAP and D3 fixed, we evaluate five normalization configurations based on Eq. (\ref{eq:norm_overview}): no processing ($none$), log compression alone ($log\_only$), and fitted path-, order-, and modulation-level normalization. 
\begin{table}[h]
    \vspace{-1.7em}
    \centering
    \small
    \caption{
        Rows and columns specify the groupings for mean centering and standard-deviation scaling, respectively, with diagonal entries corresponding to $log\_path$, $log\_order$, and $log\_modulation$. Results are reported as ``Avg. (best)''. Bold indicates the lowest EER.
    }
    \begin{tabular}{cccc}
        \toprule
        Mean $\backslash$ Std. & path & order & modulation\\
        \midrule
        path        & 4.16 (3.76) & 4.72 (3.31)  & \textbf{4.04} (3.42)\\
        order       & 5.17 (4.48) & 4.26 (3.69)  & \textbf{3.96} (3.68)\\
        modulation  & 4.20 (4.01) & 4.59 (4.16)  & \textbf{3.80} (3.33)\\
        \bottomrule
    \end{tabular}
    \vspace{-0.7em}
    \label{tab:Q2}
\end{table}
As shown in Table \ref{tab:Q2}, modulation-level scaling consistently yields the lowest average EER across all centering strategies. Notably, $none$ yields a lower average EER than the reused $log\_only$ baseline ($4.72\%$ vs. $4.98\%$), proving that log compression alone is insufficient without proper alignment. Crucially, applying modulation-level statistics to both centering and scaling achieves the lowest overall average EER of $3.80\%$, leading us to fix the $log\_modulation$ policy.
\begin{table}[h]
    \vspace{-1.7em}
    \centering
    \small
    \caption{
        Results are reported in average EER (\%), with columns evaluating alternative graph-node pooling operators.
    }
    \setlength{\tabcolsep}{4pt}
    \begin{tabular}{ccccc}
        \toprule
        Node & max & max+mean & gem+mean & mean+std \\
        \midrule
         EER (\%)   & 3.80 (3.33) & 3.86 (3.49)  & 3.53 (3.00) & 3.83 (3.31) \\
        \cmidrule(lr){2-5}
            & atten & max+atten  & Asym-A & Asym-B \\
         EER (\%)   & \textbf{3.25} (3.06) & 3.38 (2.54)  & 3.57 (2.64) & 3.58 (3.26) \\
        \bottomrule
    \end{tabular}
    \vspace{-0.7em}
    \label{tab:Q3}
\end{table}
Using modulation-level normalization, we evaluate different pooling strategies to compress the latent grid into graph nodes. As shown in Table \ref{tab:Q3}, attentive statistics achieves the lowest average EER of $3.25\%$, compared to $3.87\%$ for the baseline absmax. While this improvement introduces 19k additional parameters, it provides a favorable capacity-accuracy trade-off. Accordingly, attentive statistics are retained.

\subsection{Model Capacity and Scattering Resolution} 
With all structural modules fixed, we evaluate the remaining capacity and resolution hyperparameters. 
\begin{table}[h]
    \centering
    \small
    \caption{
        Hyperparameter optimization results across varying grid capacities and acoustic resolutions. Results are reported in EER (\%).
    }
    \begin{tabular}{ccccc}
        \toprule
        $C$ $\backslash$ $K$ & Param & $K=32$ & $K=48$ & $K=64$ \\
        \midrule
        32 & 83,195     & 4.51 (4.20)  & 3.81 (3.57)  & \textbf{3.39} (3.23) \\
        48 & 99,739     & 4.28 (3.86)  & 3.61 (3.20)  & \textbf{3.47} (3.35) \\
        64 & 119,867    & 4.47 (3.96)  & 3.97 (3.34)  & \textbf{3.25} (3.06) \\
        \midrule
        \midrule
        $J$ $\backslash$ $Q_1$ & Param & $Q_1=6$ & $Q_1=8$ & $Q_1=10$ \\
        \midrule
        6   & 119,725   & 7.06 (5.88)  & 7.51 (6.71)  & 6.95 (6.45) \\
        8   & 119,867   & \textbf{3.29} (2.92)  & \textbf{3.25} (3.06)  & \textbf{4.07} (2.97) \\
        10  & 120,009   & 5.81 (5.22)  & 6.53 (6.30)  & 6.59 (6.14) \\
        \bottomrule
    \end{tabular}
    \vspace{-1.7em}
    \label{tab:Q4}
\end{table}
First, varying the relative-time interval count $K$ and graph feature width $C$ reveals that increasing $K$ is more consistently beneficial than expanding $C$. The lowest EER is achieved at $K=C=64$, while $K=64, C=32$ serves as a lighter competitive configuration ($3.39\%$ EER, 83k parameters). Fixing $K=C=64$, we then investigate the acoustic geometry by varying the averaging scale $J$ and first-order filterbank resolution $Q_1$. Finally, comparing second-order resolutions shows that increasing from $Q_2=1$ to $Q_2=2$ expands the number of modulation bands from 6 to 13, further reducing the EER from $3.25\%$ to $2.93\%$. Since this adjustment alters modulation coverage, the improvement cannot be attributed to the minor parameter increment alone.

\subsection{Baseline Comparison and OOD Evaluation} 
Table~\ref{tab:Q6} compares WST-Graph against reproduced AASIST and AASIST-L baselines under a unified protocol. WST-Graph-Q82 achieves $2.92\%$ EER on ASVspoof2019 LA using 120k parameters, representing a $59.6\%$ reduction compared to AASIST. 
\begin{table}[h]
    \vspace{-1.7em}
    \centering
    \small
    \caption{
        Out-of-domain evaluation of reproduced AASIST baselines and our proposed configurations, reporting single-seed results.
    }
    \setlength{\tabcolsep}{3pt}
    \begin{tabular}{ccccc}
        \toprule
        System & AASIST & AASIST-L & Graph-Q81 & Graph-Q82 \\
        \cmidrule(lr){2-5}
        Param & 297k & 85k & 119k & 120k \\
        \midrule
        ITW         & 45.41   & \textbf{43.07}  & 46.07  & 44.46 \\
        ASV19LA     & \textbf{2.74}    & 3.45   & 3.07   & 2.92 \\
        ASV21LA     & 14.82   & 15.48  & 13.92  & \textbf{8.53} \\
        ASV21DF     & 19.96   & 21.25  & 21.00  & \textbf{18.15} \\
        ASV5T1      & 37.94   & 34.07  & \textbf{33.84}  & 35.55 \\
        FoR         & 27.51   & \textbf{10.46}  & 34.84  & 30.34 \\
        Codecfake   & 48.46   & 49.11  & \textbf{47.13}  & 49.26 \\
        ADD22T1     & 47.81   & 47.51  & 49.30  & \textbf{44.68} \\
        ADD22T3.2   & 38.91   & 31.56  & \textbf{31.48}  & 31.69 \\
        ADD23T1.2R1 & 52.01   & 49.80  & \textbf{44.63}  & 47.85 \\
        ADD23T1.2R2 & 43.31   & 37.55  & \textbf{34.38}  & 39.52 \\
        DFADD       & 45.69   & 38.82  & \textbf{15.23}  & 21.56 \\
        LibriSeVoc  & 38.11   & 41.06  & 38.14  & \textbf{33.58} \\
        Sonar       & 43.06   & 45.28  & \textbf{35.78}  & 39.53 \\
        \bottomrule
    \end{tabular}
    \vspace{-0.7em}
    \label{tab:Q6}
\end{table}
Across the 13 OOD datasets, WST-Graph-Q81 yields the lowest EER among all four systems on seven evaluation sets. WST-Graph demonstrates its most pronounced advantage on DFADD; across the remaining domains, it remains generally competitive without consistently large margins.

\section{Conclusion}
We introduced WST-Graph, a topology-preserving scattering front-end for graph-based SDD. Our sequential study demonstrates that performance is sensitive to temporal granularity, local-context extent, the semantic level used to calibrate WST paths, graph-node statistics, and modulation resolution. These findings highlight how representation geometry and information aggregation jointly shape detection. Future work will exploit the preserved, physically interpretable carrier-modulation coordinates for decision attribution and deepfake source tracing.

\vfill\pagebreak

\section{COMPLIANCE WITH ETHICAL STANDARDS}
This study uses publicly available datasets and does not involve new data collection from human participants. No ethical approval was required.

\section{ACKNOWLEDGMENT}
The authors have no relevant conflicts of interest to disclose. We use AI for coding assistance and polishing writing.




\bibliographystyle{IEEEbib}
\bibliography{strings,refs}

\end{document}